\documentclass[aps,prd,twocolumn,groupedaddress]{revtex4}
\usepackage{epsfig,amsmath,amssymb}
\newcommand{\sgm}{\sigma_{\rm m}}
\newcommand{\sgc}{\sigma_{\rm c}}
\newcommand{\dl}{D_L}
\newcommand{\etam}{\eta_{\rm m}}
\newcommand{\etal}{\eta_{\rm wl}}
\newcommand{\etai}{\eta_{\rm inst}}
\newcommand{\etac}{\eta_{\rm c}}
\newcommand{\nbar}{\bar{n}}
\newcommand{\dom}{\Delta \Omega}
\def \lleq {\lower0.9ex\hbox{ $\buildrel < \over \sim$} ~}
\def \ggeq {\lower0.9ex\hbox{ $\buildrel > \over \sim$} ~}
\newcommand{\ie}{{\it i.e.} }
\newcommand{\dm}{{d_m}}                                        
\newcommand{\bu}{\boldsymbol{\rm u}}                                        
\newcommand{\buT}{\boldsymbol{\rm u}^{\rm T}}                                        
\newcommand{\bg}{\boldsymbol{\rm g}}                                        
\newcommand{\bfn}{\boldsymbol{\rm f}}                                       
\newcommand{\bFn}{\boldsymbol{\rm F}}                                        
\newcommand{\bh}{\boldsymbol{\rm h}}                                       
\newcommand{\bhT}{\boldsymbol{\rm h}^{\rm T}}                                        
\newcommand{\bgT}{\boldsymbol{\rm g}^{\rm T}}                                        
\newcommand{\C}{\boldsymbol{\rm C}}                                        
\newcommand{\bS}{\boldsymbol{\rm S}}                                        
\newcommand{\half}{\frac{1}{2}}  
\newcommand{\beq}{\begin{equation}}
\newcommand{\eeq}{\end{equation}}

\begin{document}

\title{Possible use of self-calibration to reduce systematic uncertainties in
    determining distance-redshift relation via gravitational radiation
    from merging binaries}

\author{Tarun Deep Saini}
\email[]{tarun@physics.iisc.ernet.in}
\affiliation{Indian Institute of Science, Bangalore, 560 012, India} 
\author{Shiv K. Sethi}
\email[]{sethi@rri.res.in}
\affiliation{McWilliams Center for Cosmology
Department of Physics
Carnegie Mellon University
5000 Forbes Ave.
Pittsburgh, PA 15213 \footnote{on leave from Raman Research Institute, C. V. Raman Avenue, Sadashivanagar, Bangalore 560 080, India}}
\author{Varun Sahni}
\email[]{varun@iucaa.ernet.in}
\affiliation{Inter-University Centre for Astronomy and Astrophysics (IUCAA), 
Ganeshkhind, Pune 411 007, India}

\date{\today}
\begin{abstract}
By observing mergers of compact objects, future gravity wave
experiments would measure the luminosity distance to a large number of
sources to a high precision but not their redshifts. Given the
directional sensitivity of an experiment, a fraction of such sources
(gold plated -- GP) can be identified optically as single objects in
the direction of the source. We show that if an approximate
distance-redshift relation is known then it is possible to
statistically resolve those sources that have multiple galaxies in the
beam. We study the feasibility of using gold plated sources to
iteratively resolve the unresolved sources, obtain the self-calibrated
best possible distance-redshift relation and provide an analytical
expression for the accuracy achievable. We derive lower limit on the
total number of sources that is needed to achieve this accuracy
through self-calibration.  We show that this limit depends
exponentially on the beam width and give estimates for various
experimental parameters representative of future gravitational wave
experiments DECIGO and BBO.

\end{abstract}

% insert suggested PACS numbers in braces on next line
% insert suggested keywords - APS authors don't need to do this
%\keywords{}
\maketitle

Establishing the nature of dark energy is a paramount objective of
modern cosmology. A precise knowledge of cosmic distance to sources at
moderate redshifts ($z \lleq {\rm few}$) is essential for success in
this endeavor \cite{DE}. It has been
suggested that gravitational radiation from merging binaries (neutron
star (NS)-NS, NS-black hole (BH), and BH-BH) could be a `standard
siren' and a complementary means
(to standard candles and rulers) for probing cosmic expansion
\cite{schutz86,decigo,holz051,arun,sathya}.  Indeed, a
knowledge of the underlying physics of gravitational radiation from
binaries could help establish the luminosity distance to a NS-NS
binary to a precision of $2$ per cent (ignoring, for the time being,
the redshift-dependent error from gravitational lensing). In order to
serve as a cosmological probe however, the luminosity distance should
be known {\em as a function of redshift}.  Therefore, unlike other
probes of distance, the main systematic uncertainty in this case is
the identification (and redshift determination) of galaxies hosting
the binaries (see e.g. \cite{holz051,arun,sathya}) .

The space-borne gravitational wave observatory LISA \cite{lisa} is
expected to achieve an angular resolution of about $1'$ (for a
detailed discussion see \cite{holz051}). The volume bounded by this
angle (inclusive of the redshift uncertainty from luminosity distance
errors) is expected to contain roughly $30$ objects at $z \simeq 1$.
The directional sensitivity of next generation gravitational wave (GW)
observatories such as DECIGO \cite{decigo}, the Big Bang Observer
(BBO) \cite{BBO} and ASTROD \cite{Ni} is likely to be even better
($\sim$ few arc seconds), in which case, in only a small fraction of
cases would more than a single galaxy lie within the observational
beam \cite{holz09}.

The main source of uncertainty in using standard sirens to probe
cosmic expansion comes from misidentifying galaxies hosting the
standard sirens.  Clearly, the larger the number of galaxies within
the observational beam, the greater the chances for this to happen.
Given the enormous potential of using gravity wave standard sirens to
determine the nature of dark energy, it would clearly be desirable to
minimize this source of systematics. One possible approach rests in
determining an association between merging binaries and the gas in the
surrounding medium. Unique signatures of an 'afterglow' from such an
event in the electromagnetic spectrum could help in identifying the
source galaxy (e.g. \cite{bense08}). Statistical determination of
redshift using clustering properties of galaxies constitute another
possible approach \cite{macleod}. 

In this letter, we present a statistical iterative method to isolate
the source of the GW signal. This formulation assumes no prior
knowledge of the relation between the luminosity distance and redshift
(DZ relation). The method we propose can iteratively improve upon the
errors on DZ relation. Our method is briefly summarized as follows:
Beams containing a single galaxy with redshift consistent with the
expected distance-redshift relation would accurately and reliably
portray the DZ relationship. We shall call such sources `gold plated'
(GP) following \cite{holz09}. The expected number of such sources
depends crucially on directional sensitivity. If more than one galaxy
falls within the beam we propose to rule out non-sources through an
iteratively improved DZ relation. In our proposal we do not use
other means of constraining the DZ relation such as SN~Ia, since SN~Ia
systematics is likely to be far more complex than previously thought
\cite{SN_systematics}. Instead, we shall use GP sources to provide an
independent approximate starting DZ relation which we iterate upon. To
what precision this can be achieved depends upon the details of
cosmology as well the experimental parameters. In this letter we
investigate the efficacy as well as limitations of this iterative
self-calibration.

\begin{figure}
\includegraphics[width=0.45\textwidth]{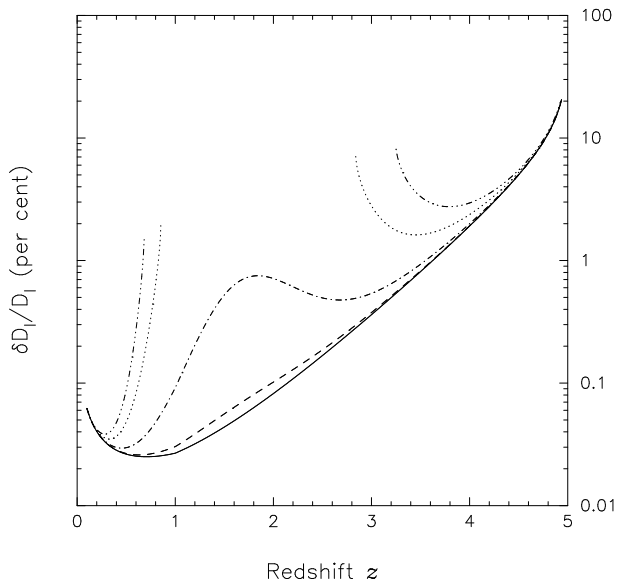}%
\caption{The maximum achievable accuracy on the DZ relation as a
function of redshift using self-calibration for the BBO case. The
solid line corresponds to perfect pointing and the dashed curve is for
the BBO pointing accuracy, where ${\dom}_{\rm BBO}$ is taken from
\cite{holz09}. The other curves correspond to degrading the pointing
accuracy to $10{\dom}_{\rm BBO}$, $50{\dom}_{\rm BBO}$,
$100{\dom}_{\rm BBO}$, corresponding to curves with increasing values
of $\delta \dl/\dl$. The last two curves have a region in the middle where
self-calibration does not work since the required number of sources
from Eq.~(\ref{eq:condition}) is larger than $3\times10^5$. The BBO
accuracy is almost as good as the case for perfect pointing, with
small departures at intermediate redshifts.}
\label{fig:fig1}
\end{figure}

\begin{figure}
\includegraphics[width=0.45\textwidth]{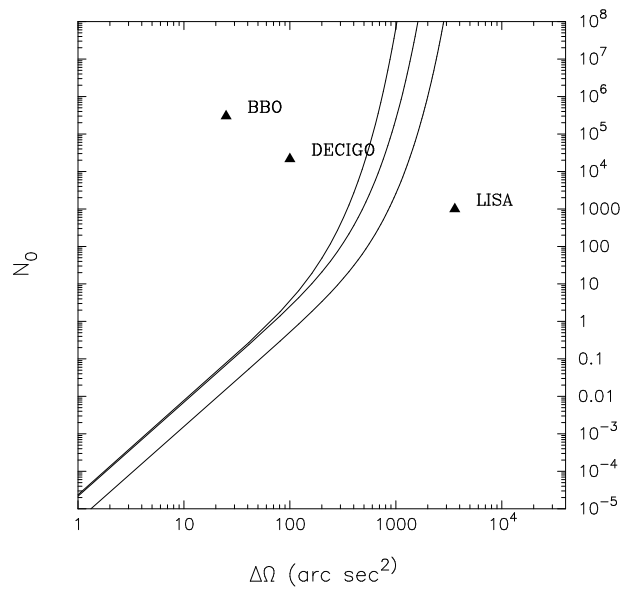}%
\caption{The minimum number of sources $N_0$ required to self-calibrate
the DZ relation at a given redshift (see Eq.~(\ref{eq:condition})) is
shown as a function of the pointing accuracy $\dom$. The curves (right
to left) correspond to $z=0.5, 1, 2$. The value of $N_0$ rises steeply
with the beam width since an increasing bin size, $\Delta z_{\rm
bin}$, necessary to suppress cosmology errors, conflicts with the
requirement of unbiased calibration. (The latter cannot be satisfied
if most GW sources are associated with multiple optical counterparts
in the pointing beam.)}
\label{fig:fig2}
\end{figure}

{\it Redshift Error-box}: The directional accuracy of a GW signal is
determined by the experimental beam width $\dom$, giving rise to the
possibility that several sources might lie within the beam. To single
out an object as the {\em unique} source of the GW signal we either
require a smoking gun signal or a criterion by which other objects
in the beam can be ruled out as possible sources. In the
presence of statistical uncertainty in DZ it is impossible to
establish precisely the redshift $z$ for a source of gravitational
radiation. One must settle instead for an uncertainty $z\pm\Delta
z_a$ where
\begin{equation}
\Delta z_a \simeq \frac{\sgm(z) \Phi(z)}{\dl} = \etam \Phi(z)\,, 
\label{eq:limit}
\end{equation}
$\Phi(z) = (d\log \dl/dz)^{-1}$ and $\sgm(z)$ is the redshift
dependent standard deviation in the luminosity distance to a single
source. The dimensionless standard deviation $\etam(z) =
\sgm(z)/\dl$ is partly due to instrumental noise and partly due
to weak lensing. The dominant uncertainty is due to lensing and
although lensing produces an asymmetric distribution of
magnifications, for our purposes we shall assume that the distribution
is described by a Gaussian with a dimensionless
standard deviation $\etal(z) = 0.042z$, derived from the results of
\cite{holz05}. In this paper, we also add in quadrature a fixed value of 
instrumental noise $\etai=0.02$ to the lensing
standard deviation to obtain $\etam^2=
\etal^2+\etai^2$. $\Delta z_a$ also contains an important {\em cosmology-dependent}
 contribution, so that the redshift error box is finally given by (see
 Appendix)
\begin{equation}
 \Delta z_a \simeq \frac{\sqrt{\sgm^2(z) +
 \sgc^2(z)}}{\dl}\Phi(z)=\etam(z)\Phi(z) \left[1 +
 \frac{\etac^2(z)}{\etam^2(z)} \right]^{1/2} \,,
\label{eq:delz}
\end{equation}
where $\sgc(z)$ is the standard deviation in DZ reflecting
uncertainty in our knowledge of the expansion history and $\etac =
\sgc/\dl$. 

Though multiple objects might lie within the beam, it may still be
possible, with a small enough value of $\Delta z_a$, to single out a
source purely on statistical grounds. However, since measurement
errors on a single source are fixed, the only way to lower the
redshift error box given by Eq.~(\ref{eq:delz}), is by reducing the
second term in that equation. 

{\it Occupation Number}: The redshift range $\Delta z_a$ together with
the beam width, $\dom$, determine the expected number of galaxies
lying within the beam that are statistically consistent both with the
approximate DZ relation as well as the measurement uncertainty. In
order to calculate the occupation number $\nbar$, defined as the mean
number of galaxies that satisfy this criterion, we have adopted the
number density of sources from \citep{holz09,kaiser,hu}. The mean
number of galaxies in the redshift range $2\Delta z_a$ turns out to be

\begin{equation}
\nbar(z) \simeq \frac{8N_{\Omega}}{h(z) \sqrt \pi} r(z)\, \exp\left[-r^4(z) \right] \dom \Delta z_a\,,
\label{eq:nbar}
\end{equation}
where we have assumed a small $\Delta z_a$ so the linear approximation
suffices. Here $r(z) = \int_0^z dz/h(z)$ is the $c/H_0$ normalized coordinate
distance, $h(z) = H(z)/H_0$ and $N_{\Omega} = 1000\,\, \rm arc\,
min^{-2}$ is the projected number density of galaxies consistent with
the Hubble Ultra Deep Field \cite{hudf}. Substituting $\Delta z_a$
from Eq.~(\ref{eq:delz}) we obtain
\begin{equation}
\nbar = \nu(z) \left[ 1 + \frac{\etac^2}{\etam^2}\right ]^{1/2}\,, 
\end{equation}
where we have defined the {\em minimum occupation number}
 $\nu(z)$ as the value of $\nbar$ when $\etac=0$, namely 
\begin{equation}
\nu(z) = \frac{8\Phi(z) N_{\Omega}}{h(z) \sqrt \pi} r(z)\, \exp\left[ -r^4(z) \right] \etam \dom\,. 
\end{equation}
We shall assume that galaxies falling within the beam are distributed
uniformly randomly, however, at the end of this letter we briefly
discuss how clustering of galaxies affects our analysis. Knowing the
occupation number $\nbar$, the probability that there be $k$ galaxies,
apart from the source galaxy, within the beam is given by $\Pr(k) =
\nbar^k
\exp(-\nbar)/k!$.  If there is only a single object in the redshift error-box then
we shall assume that it is the source of the signal.  The probability
for such instances is given by $\Pr(0) = \exp(-\nbar)$. Clearly the
limiting fraction of sources that cannot be resolved statistically is
$1-\exp(-\nu(z))$, which for $\nu(z) \ll 1$ is approximately given by
$\nu(z)$.

{\it Method}: A GP source would measure the DZ relation with an
accuracy $\etam$ at a redshift $z$. Let us consider a redshift bin,
$\Delta z_{\rm bin}$, centered at the redshift $z$. The total number
of sources in this bin is given by $\Delta N(z) = N_0f(z)\Delta z_{\rm
bin}$, where $N_0$ is the total number of GW sources at all redshifts
and $f(z)\Delta z_{\rm bin}$ is the fraction occurring in the bin
$\Delta z_{\rm bin}$. The value of $N_0$ (NS-NS binaries) for GW space
missions is expected to range from $N_0 \sim 10^3$ (LISA) to $N_0 \sim
10^6$ (BBO).  Let us assume that there are $\Delta N_{\rm GP}(z)$ gold
plated sources in the bin. These sources furnish a first estimate of
the DZ relation. Since each source has a measurement error given by
$\etam$ then clearly the zeroth error on cosmology is given by
$\eta_{\rm c0} = \etam/\sqrt{\Delta N_{\rm GP}(z)}$.

We note that if there happen to be no GP sources in the redshift bin
$\Delta z_{\rm bin}$ then $\eta_{\rm c0}$ can be obtained by fitting a
dark energy model to the GP sources at other redshifts. However, with
no a priori reason to assume a given behavior for dark energy, we
advocate this self-consistently obtained DZ relation where each
redshift is dealt with independently.

Since we now have the zeroth order information about cosmology we can
use $\eta_{\rm c0}$ to calculate the occupation number $\nbar_0 =
\nu(z)\sqrt{1+\eta^2_{\rm c0}/\etam^2}$. The zeroth knowledge of the DZ relationship
resolves some sources to give the new value of resolved sources (GP
sample and statistically resolved sources) as $\Delta N^{(1)}_{\rm
resolved}=
\Pr(0)\Delta N(z)= \Delta N(z)\exp(-\nbar_0)$, and thus provides us with
a first improved estimate $\eta_{\rm c1} = \etam/\sqrt{\Delta N(z)
\exp(-\nbar_0)}$. With this refinement in $\etac$ we can recalculate
the occupation number at the first iteration as
\begin{equation}
\nbar_1 =  \nu(z)\left [1 + \frac{1}{{\Delta N(z) \exp(-\nbar_0)}}\right ]^{1/2}\,. 
\end{equation}
It is clear that iterating further we shall obtain the recurrence
relation
\begin{equation}
\nbar_{i+1} = \nu(z)\left [1 + \frac{1}{{\Delta N(z) \exp(-\nbar_i)}}\right]^{1/2}\,.
\end{equation}
The iteration terminates when $\nbar_{i+1} = \nbar_{i}$, and therefore
the saturation cosmological uncertainty, which is the second term inside
parenthesis in the previous expression, is given by
\begin{equation}
\left(\frac{\etac}{\etam}\right)_s =  \frac{1}{\sqrt{N_0f(z)\Delta z_{\rm bin} \exp\left(-\nbar_s\right)}}\,,
\label{eq:satratio}
\end{equation}
where we substituted for $\Delta N(z)$ to explicitly show the
dependence of the saturation occupation number on the bin size and the
subscript $s$ denotes saturation value.

The uncertainty decreases as $\Delta z_{\rm bin}$ increases.  However,
since the bin size cannot be arbitrarily large, this ratio has a lower
bound, which we parameterize as $\etac^{\rm min}/\etam \equiv
\min(\etac/\etam)_s = \epsilon(z)$, where the minimum value is
obtained by choosing the largest allowed bin size. The occupation
number in this case is given by $\nbar_s =
\nu(z)\sqrt{1+\epsilon^2(z)}$, and using Eq.~(\ref{eq:satratio}) it follows 
that the bin size required to achieve this accuracy is given by
\begin{equation}
\Delta z_{\rm bin} = \frac{\exp\left[\nu(z)\sqrt{1+\epsilon^2(z)}\right]}{N_0f(z)\epsilon^2(z)}\,.
\label{eq:zbin}
\end{equation}

Averaging sources in the bin $\Delta z_{\rm bin}$ introduces a
systematic bias $\eta_c^{\rm sys}$ in the DZ relation. If the bin
size is small we can assume that the sources are distributed uniformly
in the bin. By Taylor expanding the luminosity distance $\dl$, and
taking its average over the bin we can easily obtain
\begin{equation}
\eta_c^{\rm sys} =\frac{\langle \dl(z_0) \rangle - \dl(z_0)}{\dl}  \simeq  \frac{1}{24} \frac{\dl''}{\dl}\Delta z_{\rm bin}^2\,,  
\end{equation}
which is correct up to third order in $\Delta z_{\rm bin}$, and
$\langle\,\rangle$ denotes averaging over the bin.  If we demand
$\eta_c^{\rm min} =\epsilon(z)\etam > \eta_c^{\rm sys} $ then we obtain
\begin{widetext}
\begin{equation}
\epsilon(z)\exp\left[-\frac{2}{5}\nu(z)\sqrt{1+\epsilon^2(z)}\right] \geq \frac{1}{ 24^{1/5} N_0^{2/5} f(z)^{2/5} \etam^{1/5}} \left (\frac{\dl''}{\dl}\right )^{1/5}\,,
\label{eq:accuracy}
\end{equation}
\end{widetext}
where we have substituted $\Delta z_{\rm bin}$ from Eq~(\ref{eq:zbin}).
This formula encapsulates our main result and determines the limit of
self-calibration. 

In Fig~\ref{fig:fig1} we plot $\delta D_L/D_L = \epsilon(z)\etam$ for
BBO, assuming an equality in the above expression \ie assuming that
the systematic term is equal to the random error. We have taken a flat
$\Lambda$CDM model with $\Omega_{\rm m}=0.3$ for this figure.  For
this plot we have taken $N_0$, $f(z)$ and ${\dom}_{\rm BBO}(z)$ from
\cite{holz09}.  The same figure shows the accuracy obtainable
for a degraded beam by applying a constant multiplying factor to the
BBO beam value.

{\it Necessary condition for self-calibration}: The left hand side of
Eq.~(\ref{eq:accuracy}) has a redshift dependent upper bound that we
denote as $b(z)$. Self-calibration would work only if the left hand
side is larger than the right hand side, leading to
\begin{equation}
N_0 >  \frac{1}{24^{1/2} b^{5/2}(z) f(z)\etam^{1/2}} \left(\frac{\dl''}{\dl}\right)^{1/2}\,.
\label{eq:condition}
\end{equation}
If this condition is not satisfied then the bin size required is too
large and the systematic term would dominate the random
error. Therefore, in an experiment the total number of sources $N_0$
should satisfy this inequality to self-calibrate at a given
redshift. In Fig~\ref{fig:fig2} we plot this for a few redshifts as a
function of $\dom$.

{\it Gain factor}: We now give a rough estimate of the number of
sources resolved through this method.  At a given redshift $z$ a
single source measures the luminosity distance at a fractional
accuracy $\etam$, which we can take as the cosmological accuracy
$\etac$, leading to the occupation number at the beginning to be
${\nbar}_0 = \nu(z)\sqrt{1+\etac^2/\etam^2} = \sqrt{2}\nu(z)$. To
obtain the resolved fraction at the end of iteration we shall consider
the extreme case when $N_0 \gg 1$, which along with
Eq~(\ref{eq:satratio}) gives $\epsilon_s \simeq 0$, therefore the
saturation occupation number is ${\nbar}_s \simeq
~0.414\,\nu(z)$. Since for a given occupation number $\nbar$, the fraction of
total resolved sources is given by $\exp(-\nbar)$, it is clear that
the ratio of resolved sources at the end to that at the beginning (of
iteration) is given by $\exp[0.414\,\nu(z)]$. Since $\nu(z)$ is proportional
to the pointing accuracy, the  gain is an exponential function
of the beam size. As an example, at $z \simeq 1.75$, for BBO the
minimum occupation number $\nu = 0.63$ \footnote{This number is twice
that in Fig~6 of \cite{holz09}, which we suspect is due to a
difference in interpreting $\Delta z_a$ as the allowed range. In our
analysis the allowed range for a standard deviation $\Delta z_a$ is
twice the standard deviation}, giving a maximum gain factor of $\sim
1.3$, while for DECIGO, assuming a beam linear size about a factor three
worse, the gain factor is about $\sim 11$, showing the extreme
sensitivity of gain to the directional sensitivity.

{\it Effect of clustering}: In the discussion so far we have neglected the 
impact of galaxy clustering. The effect of the galaxy
clustering can be taken into account by replacing $\nbar$ with
\begin{equation}
\nbar\left (1+ \frac{1}{\Delta V} \int \xi dV \right).
\label{eq:clus}
\end{equation}
Here $\Delta V$ is the volume bounding the redshift and angular
error-box ($\Delta z$ and $\Delta \Omega$) in the determination of the
source. $\xi$ is the two-point correlation function of galaxy
clustering and the integral extends over $\Delta V$. 

Here we give estimates of the impact of clustering (second term of
Eq.~(\ref{eq:clus})) for BBO and DECIGO configurations at $z \simeq
2$. In the approximation, valid for these cases, in which  the
(comoving) linear size corresponding to angular resolution $l_\perp$
is much smaller than the distance corresponding to the radial distance
$l_{\parallel}$ of the (minimum i.e. when $\etac \rightarrow 0$) redshift error, one can readily show
that
\begin{equation}
\frac{\Delta \nbar}{\nbar} =  \frac{1}{\Delta V}\int \xi dV \simeq  \left( {l_\perp \over 0.2  \,  {\rm Mpc}} \right )^{-0.8} \left( {l_{\parallel} \over 300 \,  {\rm Mpc}} \right )^{-1}.
\label{eq:clus1}
\end{equation}
$\Delta \nbar/\nbar \simeq \{1, 0.45 \}$ for the BBO and the DECIGO at
$z\simeq 2$. Eq.~(\ref{eq:clus1}) shows that this term scales
inversely with $l_{\parallel}$, and therefore the effect of clustering
would be less important in the beginning of the iteration process when
$\etac$ could be appreciable but would be increasingly important as
the maximum achievable precision is approached.  

We have shown that by iterating over a self-consistently obtained DZ
relation from resolved gravity wave sources it is possible to improve
the luminosity distance (DZ) relation and therefore isolate those
sources that initially are unidentifiable (owing to multiple objects
in the pointing beam).

However, due to the fact that in this process only the cosmological
errors are reduced, the limiting resolved set crucially depends on the
pointing accuracy at a given redshift.  We have derived analytical
expressions for the final accuracy reached on the DZ relationship as
well as the condition for successful self-calibration (Eq~\ref{eq:condition} and
Figure~2). Our formulation will help future GW probes grapple with the
issue of redshift measurement uncertainty due to the presence of
multiple objects within their beam (Figure~1).  A comprehensive
analysis using simulated data to estimate cosmological constraints
arising from future GW experiments will be presented in a companion
paper.

\begin{appendix}
\section{}

We now derive the probability distribution function (PDF) for the
source redshift given a cosmological model and a distance measurement
to a GW source. We first derive the general formula and then specialize
to the local approximation used in this paper. 

Let the measured distance be given by $\dm$. To quantify cosmology
errors we employ a linear model for the DZ relation,
\beq
\dl(z,\bh) = \sum_i^N h_i f_i(z) = \bhT \bfn\,,
\eeq
where $\bh$ are the $N$ parameters of the model, $f_i$ are $N$
arbitrary functions of redshift, and we have defined $\bfn =
[f_1(z),f_2(z),\ldots,f_N(z)]$. The unknown redshift is to be treated
as a parameter of the model. The simplest choice is $f_i = z^{i-1}$,
leading to a polynomial form for $\dl(z)$. The parameters $\bh$ are not
known precisely and are described by the Gaussian distribution
\beq
P(\bh) = \frac{1}{(2\pi)^{N/2} \sqrt{\det{\C}}}\exp\left [-\half (\bhT-\bhT_0) {\C}^{-1} (\bh-{\bh}_0) \right]
\label{eq:parmcosmo}
\eeq
where ${\C}$ is the covariance matrix, obtained by fitting the model
to the resolved sources (or to other data sets), and $\bh_0$ are the
best fit parameters. We employ bold lower case letters to denote
column matrices and bold capital letters to denote second rank
matrices. Employing the Bayes theorem we can write down the posterior
probability for the parameters of the model as
\beq
P(z,\bh|\dm) \propto P(\dm| z,\bh) P(\bh) P(z)\,,
\label{eq:zdist}
\eeq
where $ P(\bh)$ is the prior PDF for the parameters
$\bh$ given by Eq.~\ref{eq:parmcosmo}, the prior $P(z)$ is
assumed to be flat, and 
\beq
P(\dm| z,\bh) = \frac{1}{\sqrt{2\pi}\sigma_m}\exp\left [- \frac{(\dm - \bhT\bfn)^2}{2\sigma_m^2} \right ]\,\,.
\eeq
The PDF in Eq.~\ref{eq:zdist} is a function of redshift $z$ and $\bh$,
therefore the posterior PDF for the source redshift can be obtained by
integrating over $\bh$ 
\beq
P(z|\dm) \propto \int  P(\dm| z,\bh) P(\bh) \,d^Nh\,,
\eeq
which can be expressed through variables $\bg=\bh-{\bh}_0$ and $\chi = \dm -
\bhT_0\bfn$ as
\begin{widetext}
\beq
P(z|\dm) \propto \int \exp\left [ - \frac{\chi^2  - 2\chi \bgT\bfn + \bgT \bFn \bg}{2\sigma_m^2} - \half \bgT {\C}^{-1} \bg \right ] d^Ng\,, 
\eeq
\end{widetext}
where the coefficients that do not contain $z$ and $\bg$ have been
dropped, and we have defined a second rank matrix $\bFn = \bfn
\otimes \bfn$. If we define a matrix $\bS = {\C}^{-1} +
\bFn/\sigma_m^2$ then this equation takes the from
\beq
P(z|\dm) \propto \int \exp\left [ - \frac{\chi^2 - 2\chi \bgT\bfn}{2\sigma_m^2} - \half \bgT {\bS} \bg \right ] d^Ng\,. 
\eeq
Translating the coordinate system in the parameter space $\bg = \bu +
{\bu}_0$, where ${\bu}_0$ is such that $\bS {\bu}_0 = \chi
\bfn/{\sigma_m^2}$, we finally obtain
%\begin{widetext}
%\beq
%P(z|\dm) \propto P(z)\int \exp\left[-\frac{\chi^2}{2\sigma_m^2} + \half \left(\frac{2\chi}{\sigma_m^2} (\buT + {\bu}_0^{\rm T}) \bfn  - (\buT {\bS} \bu + 2\buT \bS \bu_0 + {\buT}_0 \bS {\bu}_0 \right)\right ] d^Nu\,. 
%\label{eq:shift}
%\eeq
%\end{widetext}
\beq
P(z|\dm) \propto \int \exp\left[ \half \left( -\frac{\chi^2}{\sigma_m^2} + \frac{\chi}{\sigma_m^2} {\bu}_0^{\rm T}\bfn -\buT {\bS} \bu  \right)\right ] d^Nu 
\eeq
Carrying out the integration and dropping all terms that do not depend on the redshift we obtain
\beq
P(z|\dm) \propto \det(\bS) \exp\left[-\frac{\chi^2}{2\sigma_m^2} \left (1- \frac{\bfn^{\rm T} {\bS}^{-1} \bfn}{\sigma_m^2}\right)  \right]
\eeq
Recalling that  $\chi = \dm -
\bhT_0\bfn$,  we find that the redshift probability distribution is
centered at the redshift predicted for the distance $\dm$ by the best
fit model $d(z;{\bh}_0)$. Since the functions $\bfn$ are redshift
dependent, the precise behavior of this function is complicated.
$P(z|\dm)$ can be normalized in the range $z=0$ to $z=z_{\rm max}$, and
would, in general, produce an asymmetric distribution, due to the
manner in which the cosmological errors scale with redshift.

{\it Local approximation:} Since the PDF peaks at $\chi=0$, we can
define a redshift $z_0$ through $ \dm = \bhT_0\bfn(z_0)$. If the
cosmology is determined precisely then we can assume the redshift PDF to
decline rapidly away from $z_0$, and therefore we can
replace $\bhT_0\bfn(z) = \bhT_0\bfn(z_0) + \bhT_0\bfn'(z_0) (z-z_0)$,
implying $\chi = -\bhT_0\bfn'(z_0) (z-z_0)$. Then, at the same level
of accuracy we can replace $\bfn \equiv \bfn_0 = \bfn(z_0)$ in
$\bfn^{\rm T} {\bS}^{-1} \bfn$. To evaluate $\bfn^{\rm
T}_0{\bS_0}^{-1} \bfn_0$, we need an expression for
${\bS_0}^{-1}$. Noting that $\bFn_0 $ is a rank one matrix, we have \cite{mil81}
\begin{equation}
{\bS_0}^{-1} = \C  - \frac{1}{1+g} \frac{\C \bFn_0 \C}{\sigma_m^2}\,,
\end{equation}
where $g = {\rm tr}\, \bFn_0 \C/\sigma_m^2$. Noting that $\sigma_c^2 =
{\rm tr}\, \bFn_0 \C$, it can be readily shown that
\begin{equation}
\bfn_0 {\bS_0}^{-1} \bfn_0 = \frac{ \sigma_m^2 \sigma_c^2} { \sigma_m^2 + \sigma_c^2} 
\end{equation}
The redshift probability distribution function can be now written explicitly as 
\beq
P(z|\dm)  = \frac{1}{\sqrt{2\pi} \sigma^2_z} \exp\left[-\frac{(z-z_0)^2}{2\sigma_z^2}   \right]
\eeq
where $\sigma_z = \sqrt{(\sigma_m^2 + \sigma_c^2)}/D'_L(z_0)$.

\end{appendix}

\bibliography{}

\end{document}